\documentclass[%
reprint,
superscriptaddress,
onecolumn,              
showkeys,               
 amsmath,amssymb,
 aps,
]{revtex4-2}

\usepackage{graphicx}
\graphicspath{{./figs/}}
\usepackage{subcaption}
\usepackage[colorlinks=true, linkcolor=blue, citecolor=blue, urlcolor=blue]{hyperref}

\begin{document}

\preprint{APS/123-QED}

\title{Alignment-induced depression and shear thinning of anisotropic granular media}

\author{Huzaif Rahim  }
  \email{huzaif.rahim@fau.de}
\affiliation{%
{
Institute for Multiscale Simulation, \\ Friedrich-Alexander-Universit\"at Erlangen-N\"urnberg, \\ Cauerstrasse 3, 91058 Erlangen, Germany
}
}%

\author{Vasileios Angelidakis}%
\affiliation{%
{
School of Natural and Built Environment, \\ Queen's University Belfast, \\ David Keir Building, BT9 5AG Belfast, United Kingdom
}
}%

\author{Thorsten P\"oschel}%
\affiliation{%
{
Institute for Multiscale Simulation, \\ Friedrich-Alexander-Universit\"at Erlangen-N\"urnberg, \\ Cauerstrasse 3, 91058 Erlangen, Germany
}
}%

\author{Sudeshna Roy}%
\affiliation{%
{
Institute for Multiscale Simulation, \\ Friedrich-Alexander-Universit\"at Erlangen-N\"urnberg, \\ Cauerstrasse 3, 91058 Erlangen, Germany
}
}%

\begin{abstract} 
When granular materials of shape-anisotropic grains are sheared in a split-bottom shear cell, a localized shear band is formed with a depression at its center. This effect is closely related to the alignment of the particles with aspect ratio (AR), which, in turn, influences the local packing density, the stress distribution, and the system's overall bulk rheology. Particles with large AR tend to align with the shear direction, which increases the packing density in the shear band and affects rheological properties like stress, macroscopic friction coefficient, and effective viscosity. A scaling law  correlates particle AR to macroscopic friction and effective viscosity, revealing shear-thinning behavior
in bulk and near the surface.

\end{abstract}

\maketitle

\section{Introduction}
Granular materials, consisting of shape-anisotropic particles such as elongated grains, are used in a wide range of applications, including the food industries, chemical industry, and geology \cite{aranson2006patterns, torres2017flowability, cambou2004anisotropy}. Shearing, pouring, or shaking of such particles induce local or global orientational orders, resulting in anisotropic microstructures that influence material stress behavior and flow characteristics \cite{nagy2023flow, pol2022kinematics, borzsonyi2012orientational, borzsonyi2013granular, campbell2011elastic, guo2013granular, wegner2014effects}. 
Specifically, shearing in a split-bottom shear cell causes such elongated grains to accumulate in the center, forming a heap \cite{fischer2016heaping, stannarius2017heaping, wortel2015heaping}. Wortel et al. first reported this phenomenon, attributing it to a secondary convective flow driven by an out-of-plane misalignment between particle orientation and flow streamline \cite{wortel2015heaping}. A significant elevation in the surface profile, approximately 30\% higher than the initial filling height, was observed at the central region of the shear cell, resulting from secondary flows induced by shearing prolate and oblate grain shapes \cite{fischer2016heaping}. Concurrently, This elevation corresponds to a depression observed on the shear band surface, which is absent in the flow of spherical particles\cite{wortel2015heaping, fischer2016heaping}. The secondary flow in granular media has been studied in Couette cell and split-bottom shear cell with densely packed spherical particles \cite{dsouza2021dilatancy, dsouza2017secondary, krishnaraj2016dilation}, and in nonspherical isotropic grain shapes \cite{mohammadi2022secondary}. While research has focused on the influence of particle shape, alignment, and secondary flows, the underlying mechanisms for changes in the shear band surface profile of anisotropic particles in the split-bottom shear cell remain elusive.

When the stress exceeds the yield threshold, granular materials flow like non-Newtonian fluids  \cite{de2005non, campbell2006granular}. Specifically, their viscosity, $\eta = \tau/\dot{\gamma}$, decreases as the shear rate $\dot\gamma$ increases, a phenomenon known as shear-thinning \cite{roy2017general}. This effect is quantified by the inertial number, $I$, where viscosity is inversely related to the strain rate, $\tau/\dot{\gamma}\propto I^{-\alpha}$, with $\alpha = -1$ or -2, depending on the local confining pressure \cite{luding2007effect}. The shear-thinning behavior is influenced by the particle shape, the particle interactions, and gravity \cite{roy2017general, hosseinpoor2021rheo,marteau2021experimental}. Despite its importance, by now the effect of the particle shape on the viscosity and shear-thinning in granular flows is not well known.

Cylindrical split-bottom shear cells have been used to study the effects of particle shape on granular behavior, such as dilatancy, orientational order, secondary flows, and stress distribution \cite{dijksman2010granular,fenistein2003wide,fenistein2004universal, wegner2014effects, borzsonyi2013granular, fischer2016heaping, wortel2015heaping, mehandia2012anomalous, depken2007stresses}. Expanding on this approach, subsequent studies have adopted linear split-bottom shear cells (LSC), as demonstrated in \cite{ries2007shear,dsouza2021dilatancy,roy2021drift}. These configurations have similarities with cylindrical ones and offer the advantage of periodic boundary conditions along the flow direction. 


We use the LSC to investigate how particles with different aspect ratios (AR) form a depression on the shear band surface. We analyze the flow dynamics of elongated particles, focusing on understanding their behavior from both micro and macro perspectives. We also explore the rheology of non-Newtonian granular flows with anisotropic particles, drawing parallels with stress responses observed with secondary flows in non-Newtonian fluids, known as the \emph{Weissenberg effect}. Our results suggest a scaling law for the effective viscosity of anisotropic particles.

\section{Numerical setup and materials}
\begin{figure*}[htb]
\centering
\begin{subfigure}{0.48\linewidth}
    \includegraphics[trim={0cm 0cm 0cm 0cm},clip,width=\linewidth]{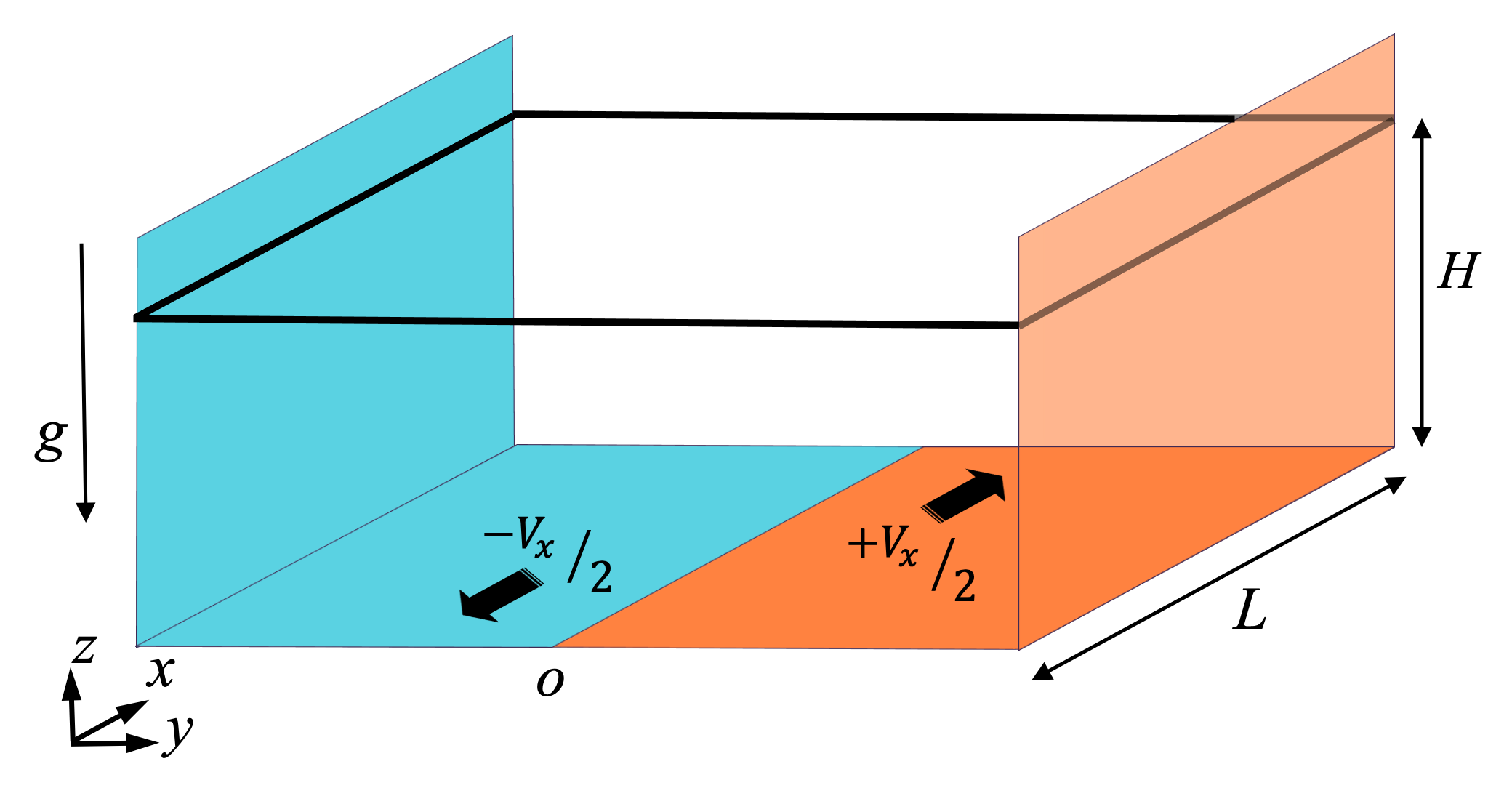}
    \subcaption{}
\end{subfigure}
\hfill
\begin{subfigure}{0.48\linewidth}
    \includegraphics[trim={0cm 6cm 0cm 0cm},clip,width=\linewidth]{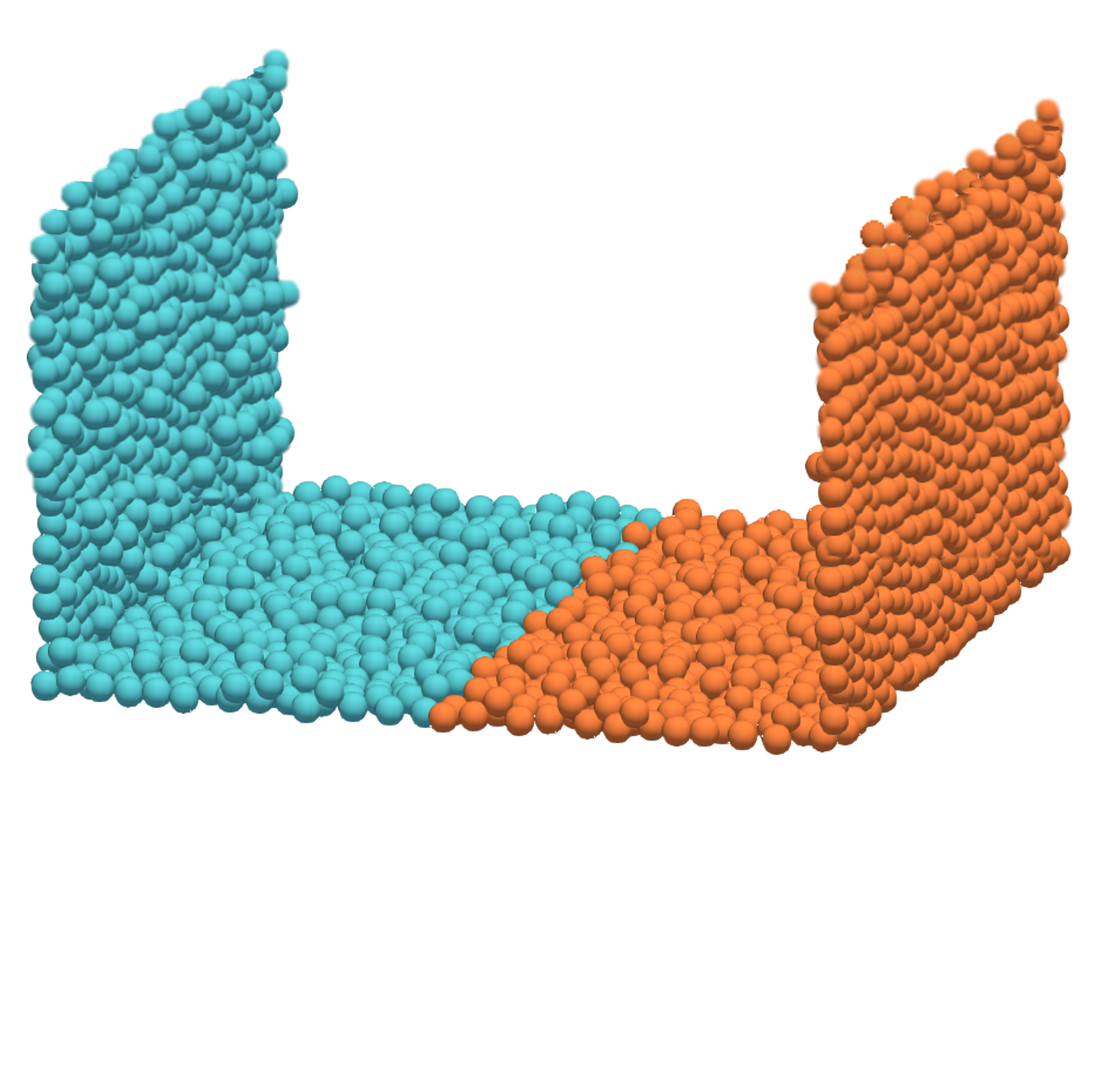}
    \subcaption{}
\end{subfigure}
\begin{subfigure}{0.48\linewidth}
    \includegraphics[trim={0cm 0cm 0cm 0cm},clip,width=\linewidth]{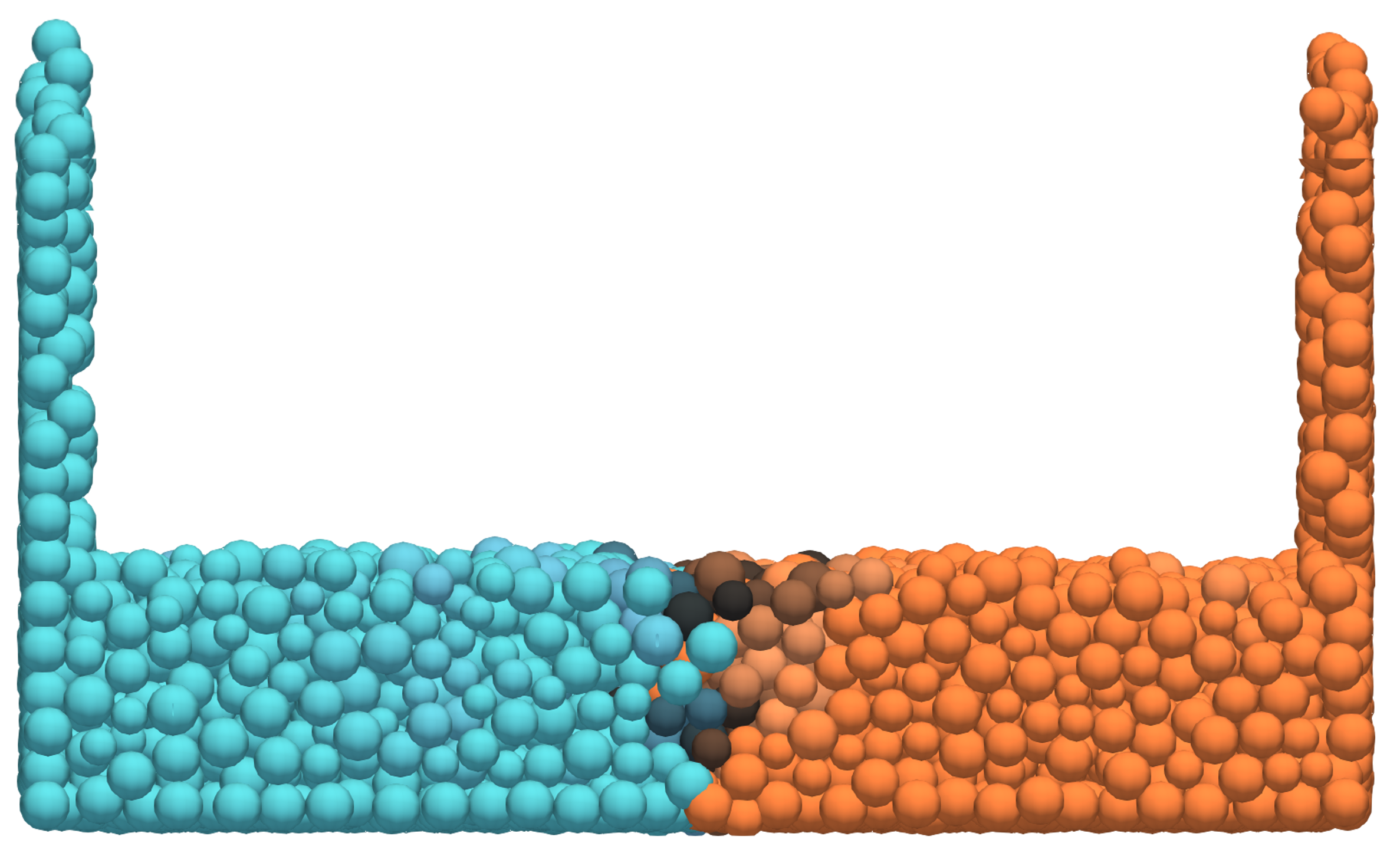}
    \subcaption{}
\end{subfigure}
\hfill
\begin{subfigure}{0.48\linewidth}
    \includegraphics[trim={0cm 0cm 0cm 0cm},clip,width=\linewidth]{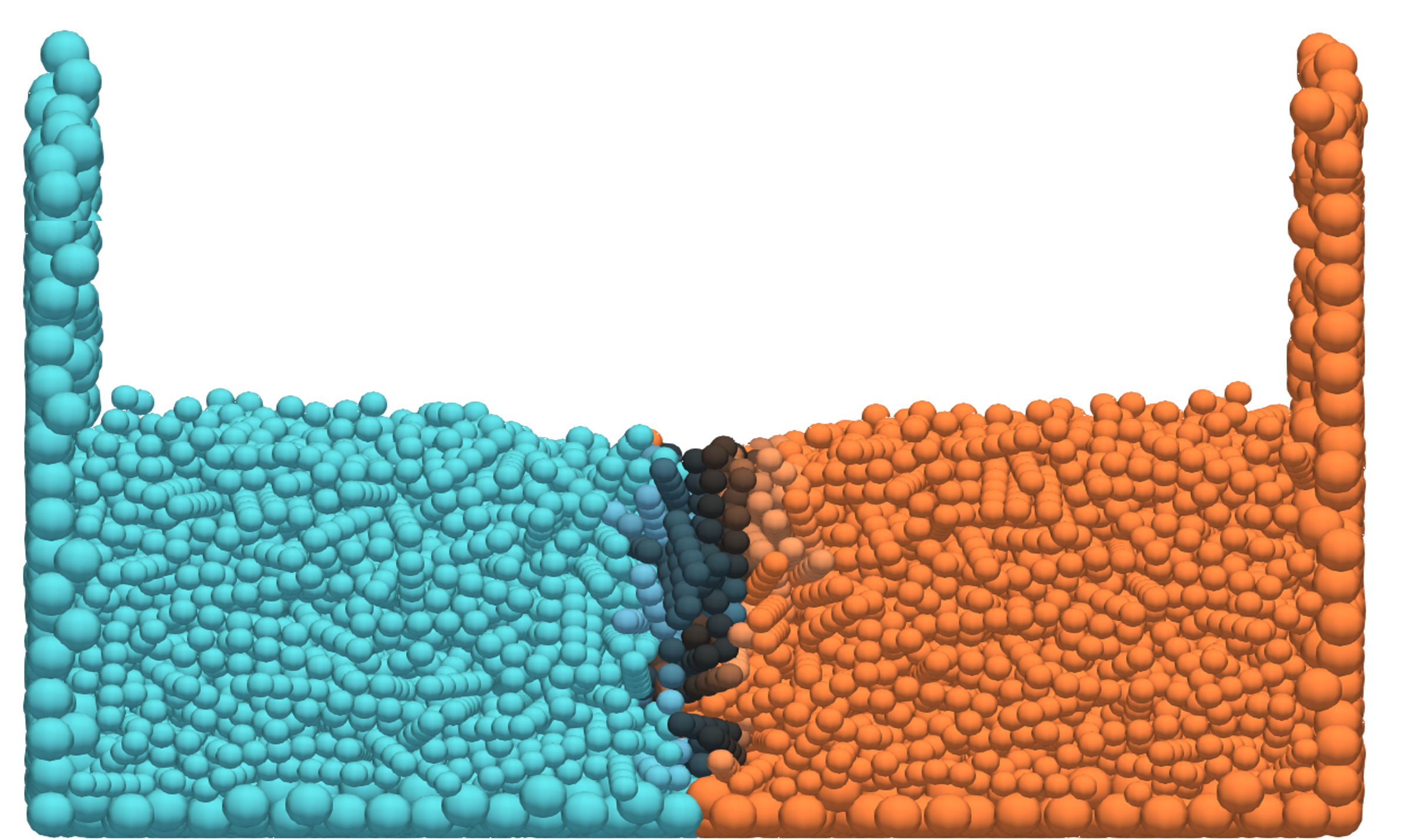}
    \subcaption{}
\end{subfigure}

\begin{subfigure}{0.70\linewidth}
    \includegraphics[trim={0cm 0cm 0cm 0cm},clip,width=\linewidth]{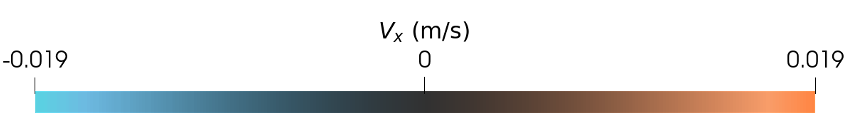}
\end{subfigure}

    \caption{(a) Sketch of the linear split-bottom shear cells (LSC), where `O' represents the split position, $V_{x}$ is the shear velocity, and $g$ is the acceleration due to gravity. (b) Spherical particles constitute the walls of the shear cell. The shear cell is filled with (c) spherical particles, showing a flat surface, and (d) elongated particles, showing a depression on the surface of the shear band.}
    \label{fig:LSC}
\end{figure*}

The LSC consists of two L-shaped walls separated by a split through the origin `O', sliding past each other, as shown in \autoref{fig:LSC} (a). The space between the walls is filled with particles, deposited under gravity, and then sheared. The left and right L-shaped walls move in opposite directions at velocities $-V_{x}/{2}$ and $V_{x}/{2}$, respectively, where $V_{x} = 0.038\,\text{m/s}$ is the shear velocity. The width, $L$, of the box, the filling height, $H$, and the coordinates are defined in \autoref{fig:LSC}(a): The $x$-direction aligns with the flow, the $y$-direction is lateral to it, and the $z$-direction 1s orthogonal to the bottom wall. The gravitational force acts downward in the negative $z$-direction. The walls are made of spherical particles of diameter $d_{p} =8.55\,\text{mm}$ with a volume equal to that of the bulk particles, as shown in \autoref{fig:LSC}(b). The shear cell dimensions are $25 d_{p}$, and periodic boundary conditions are applied in the $x$-direction. In the steady-state condition, a shear band emerges from the split position, extending laterally and upwards as shown by the regions of velocity gradient in \autoref{fig:LSC}(c) and (d).

\subsection{Contact model and material parameters}
We use the visco-elastic Hertz-Mindlin contact model \cite{thornton2015granular, thornton2013investigation} to calculate the normal and tangential elastic contact forces between particles. The model defines the normal force as
\begin{equation}
    \vec{F_n} = \min\left(0, -\rho \xi^{3/2} - \frac{3}{2}A_n\rho\sqrt{\xi}\dot{\xi}\right) \vec{e_n}\,,
\end{equation}
where $\xi = R_i + R_j - \vert\vec{r}_i-\vec{r}_j\vert$ is the compression of two interacting particles $i$, $j$ of radii $R_i$ and $R_j$ at positions $\vec{r}_i$ and $\vec{r}_j$, $\vec{e}_n = (\vec{r}_i-\vec{r}_j)/\vert\vec{r}_i-\vec{r}_j\vert$ is the normal unit vector, $A_n = 3\times 10^{-4}$s is the normal dissipative parameter, calculated as in \cite{muller2011collision}. The effective stiffness of the Hertzian contact model is
\begin{equation}
    \rho = \frac{4}{3} \, E^* \, \sqrt{R^*} 
    \label{eq:Hertz_rho}
\end{equation}
where $E^*$ is the effective elastic modulus, and $R^*$ is the effective radius. The effective elastic modulus $E^*$ is given by
\begin{equation}
     E^* = {\bigg(\frac{1-\nu_i^2}{E_i} + \frac{1-\nu_j^2}{E_j}\bigg)}^{-1}\,,
\end{equation}
where $E_{i}$ is the elastic modulus and $\nu_{i}$ is the Poisson's ratio of the material of particle $i$. 

We model the tangential viscoelastic forces using the no-slip expression by Mindlin \cite{mindlin1949compliance} for the elastic part and Parteli and P\"oschel \cite{parteli2016particle} for the tangential dissipative constant $A_t \approx 2 A_n E^*$, which are capped by the static friction force.
The tangential force is given by
\begin{equation}
    \vec{F_t} = -\min \left[ \mu|\vec{F_n}|,  \int 8 G^{*}\sqrt{R^* \xi} \,ds 
    + A_t  \sqrt{R^* \xi} v_t \right] \vec{e_t}\,,
\end{equation}
where $\mu$ is the friction coefficient, $G^*$ is the effective shear modulus calculated as $G^*=\left(\frac{2-\nu_i}{G_i} + \frac{2-\nu_j}{G_j}\right)^{-1}$, which for two identical materials in contact simplifies to $G^*=\frac{G}{2\left(2-\nu\right)}$ and $\,ds$ is the tangential relative displacement of the particles.

The material parameters are given in \autoref{tab:material_parameters}.
\begin{table}[htb!]
\caption{DEM simulation parameters.}
\label{tab:material_parameters}
\begin{center}
\begin{tabular}{l@{\quad}l@{\quad}ll}
\hline
\multicolumn{1}{l}{\rule{0pt}{18pt}
                   variable}&\multicolumn{1}{l}{unit}&{value}\\[2pt]
                   \hline\rule{0pt}{12pt}\noindent
                   elastic modulus ($E$)  & MPa& 10\\
                   sliding friction coefficient ($\mu$)  & -& 0.30\\
                   Poisson's ratio ($\nu$)  &
                   -& 0.35\\[2pt]
                   particle density ($\rho$)  & kg/m$^3$& 850\\[2pt]
\hline                
\end{tabular}
\end{center}
\end{table}

For the material density and friction coefficient, we used the values of wooden pegs \cite{fischer2016heaping}. The reduced Young's modulus was chosen to be smaller to keep the computer time feasible.

\section{Particle shapes}
\begin{figure*}[htbp]
    \centering
    \includegraphics[trim={0cm 0cm 0cm 0cm},clip,width=0.8\linewidth]{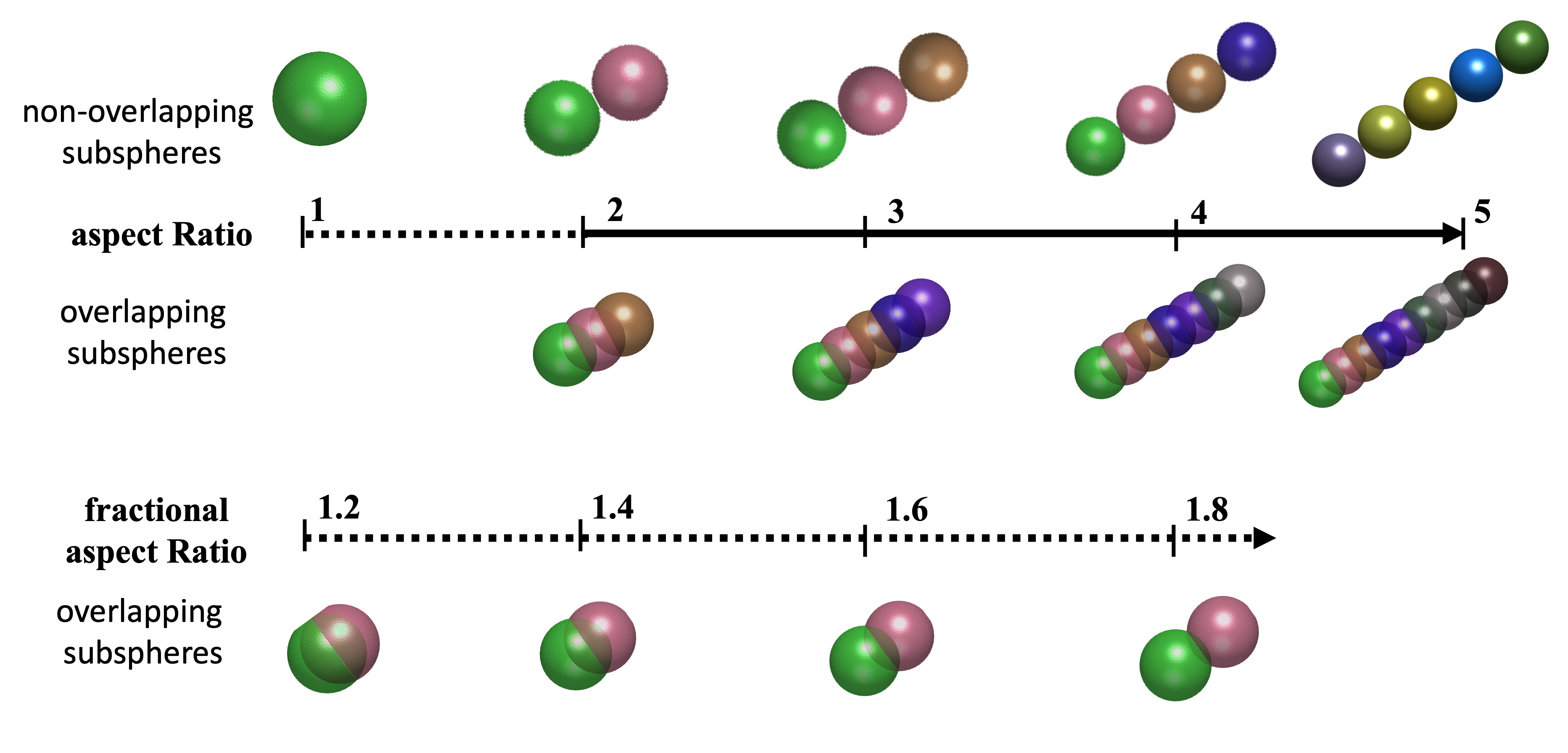}
    \caption{Multisphere particles for aspect ratio $\text{AR}\in (1,5)$ with both overlapping and non-overlapping subspheres.}
    \label{fig:particles}
\end{figure*}
We perform numerical simulations by solving Newton's equation of motion in the LSC. The particle AR is varied in the interval $[1,5]$. Contact detection of particles with nonspherical shapes is numerically difficult. To overcome this issue, these particles are approximated by merging multiple spheres (subspheres) into a single nonspherical particle \cite{abou2004three,kodam2009force,cabiscol2018calibration}. We use particles with different AR, represented by both non-overlapping and overlapping subspheres \cite{kloss2012models, pei2015numerical}. For instance, the former correlates the number of spheres with AR, while the latter uses 3, 5, 7, and 9 spheres to represent AR$=2,\dots 5$, respectively. We also study particles with non-integer AR, such as 1.2, 1.4, 1.6, and 1.8, using overlapping subspheres. Details of these particle configurations are shown in \autoref{fig:particles}. We simulate 4318 particles for each AR. All particles have a uniform volume to maintain consistent hydrostatic pressure across all simulations.

\section{Micro-macro transition}
To extract the macroscopic fields, we used a coarse-graining technique to obtain the macro-parameters through precise calculation of sphere overlap volumes and mesh elements, as outlined by Strobl et al. \cite{strobl2016exact}. We simulated the system for 100\,s real-time. To obtain steady-state data, we average over the $x$-direction and over time from 80 to 100\,s, generating continuum fields as $Q(y, z)$.  
We focus on the regime beyond the critical state, where the material deforms continuously at a constant strain rate without further stress changes. This is typically observed at the shear band's center. We define the shear band region as the region where the local strain rate at height $z$ exceeds the critical strain rate, $\dot{\gamma}_{c(z)} \equiv 0.8\dot{\gamma}_{\text{max}}(z)$, where $\dot{\gamma}_{\text{max}}(z)$ is the maximal value at the given height $z$.

\section{Results and discussions}

\subsection{Shear-induced particle alignment}
The aim of our investigations is to clarify the origin of the depression in the spatial proximity of the shear band. We start by analyzing the particles' alignment with the shear direction. The alignment angle $\theta_{x}$ measures the orientation of the principal vector of an elongated particle, $\Vec{P_{v}}$, relative to the shear direction, $\Vec{X}$:
\begin{equation}\label{dotprod}
   \theta_{x} \equiv \frac{\pi}2-\left|\frac{\pi}2- \arccos \left(\frac{\Vec{P_{v}} \cdot \Vec{X}}{|\Vec{P_{v}}||\Vec{X}|}\right)\right|\,.
\end{equation}
The limits, $\theta_{x}=0$ and $\theta_{x}=\pi/2$, indicate perfect alignment and misalignment, respectively. 

\begin{figure*}[htb!]
\centering
\begin{subfigure}{0.48\linewidth}
    \includegraphics[trim={0cm 0cm 0cm 0cm},clip,width=\linewidth]{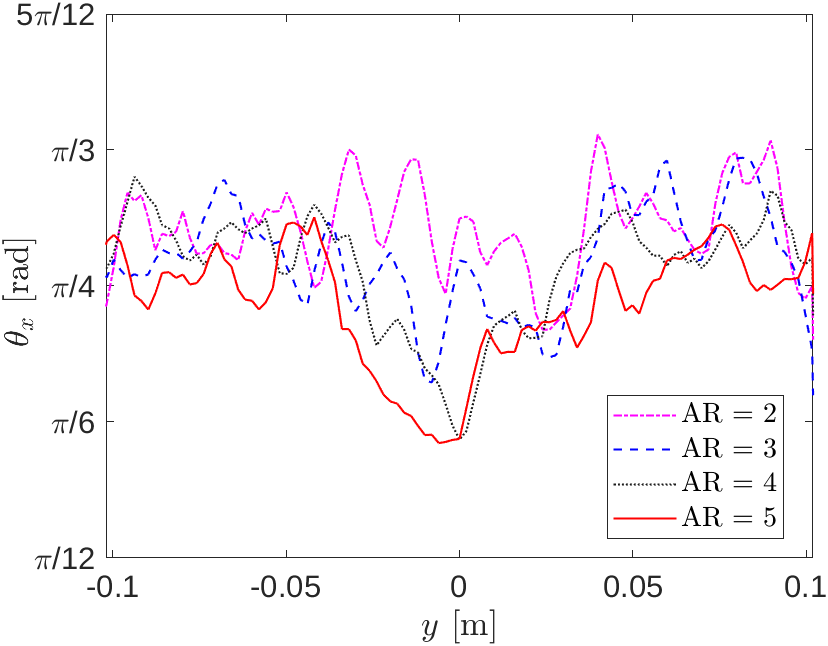}
    \subcaption{}
\end{subfigure}
\hfill
\begin{subfigure}{0.48\linewidth}
    \includegraphics[trim={0cm 0cm 0cm 0cm},clip,width=\linewidth]{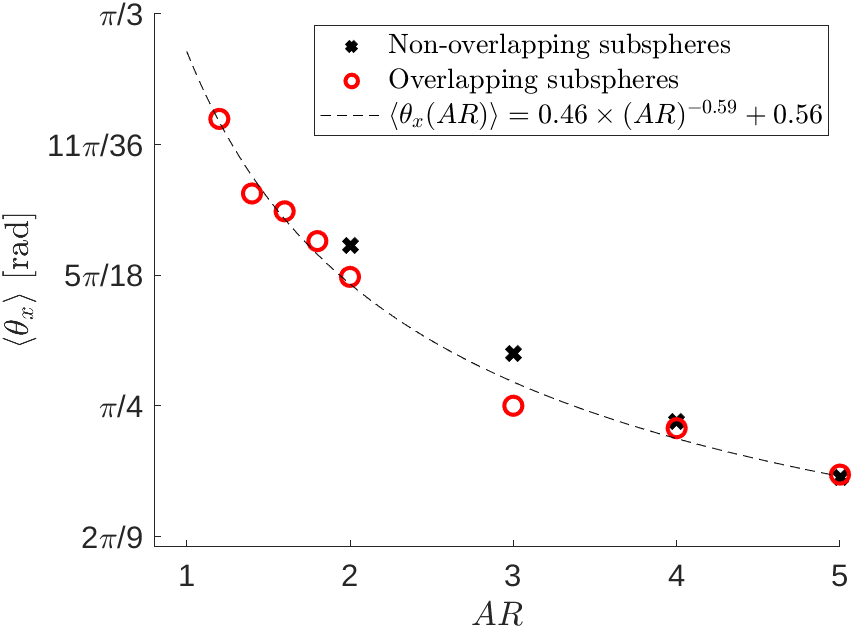}
    \subcaption{}
\end{subfigure}
\caption{(a) Particle alignment angle $\theta_{x}$ as a function of the $y$-position. (b) Global average orientation angle $\left<\theta_x\right>$ as a function of the AR for particles with non-overlapping and overlapping subspheres. The dashed line shows a fit.}
\label{fig:orientation} 
\end{figure*}

We explored the relation between the particle alignment and AR of elongated particles by averaging along the LSC's $x$ and $z$ directions. The correlation between $\theta_{x}$ and the particle's position in the $y$ direction is shown in \autoref{fig:orientation}(a). Particles with a smaller AR maintain a uniform,  $\theta_{x}$ across all $y$ positions. However, for particles with larger AR (3,4,5) we find a clear decrease in $\theta_{x}$ in the vicinity of the shear band. For instance, for AR=5, $\theta_{x}$ inside the shear band is 44\% less ($\theta_{x} \approx 0.15 \pi$) compared to outside the shear band where $\theta_{x} \approx 0.27 \pi$. This indicates that elongated particles align prominently with the shear direction inside the shear band, emphasizing the shear band's key role in the orientational ordering of particles.

The global average orientation of the particles with the shear direction, defined by 
\begin{equation}\label{OrientationAvg}
\left<\theta_{x}\right> \equiv \frac{1}{N} \sum_{i=1}^{N} \theta_{x,i} \,.    
\end{equation}
decreases with increasing AR for particles regardless of their subsphere configuration, see \autoref{fig:orientation}(b), in agreement with earlier studies \cite{borzsonyi2013granular}. 

\subsection{Packing density}
\begin{figure*}[htb!]
\centering
\begin{subfigure}{0.48\linewidth}
    \includegraphics[trim={0cm 0cm 1.6cm 0cm},clip,width=\linewidth]{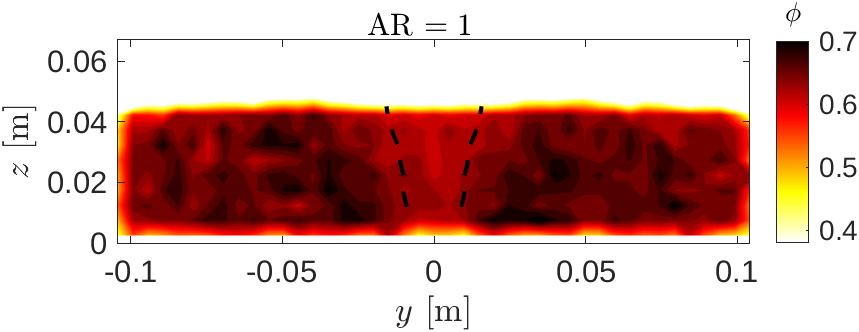}
    \subcaption{}
\end{subfigure}
\hfill
\begin{subfigure}{0.48\linewidth}
    \includegraphics[trim={0cm 0cm 1.6cm 0cm},clip,width=\linewidth]{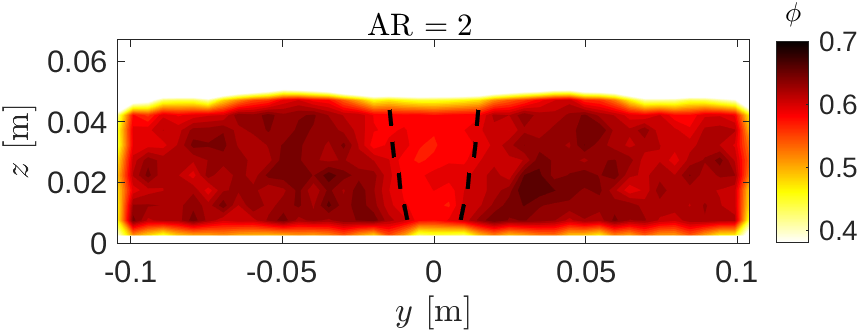}
    \subcaption{}
\end{subfigure}
\begin{subfigure}{0.48\linewidth}
    \includegraphics[trim={0cm 0cm 1.6cm 0cm},clip,width=\linewidth]{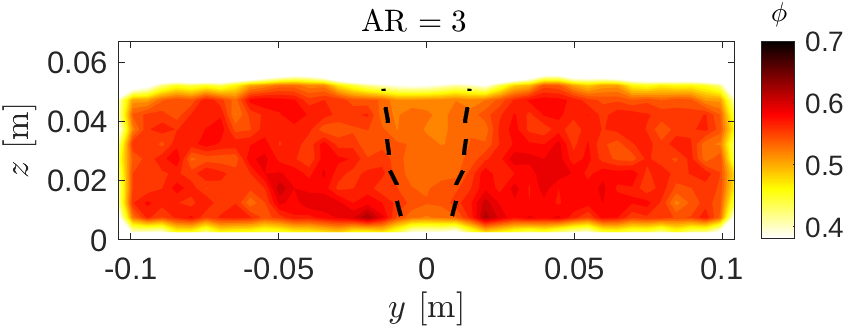}
    \subcaption{}
\end{subfigure}
\hfill
\begin{subfigure}{0.48\linewidth}
    \includegraphics[trim={0cm 0cm 1.6cm 0cm},clip,width=\linewidth]{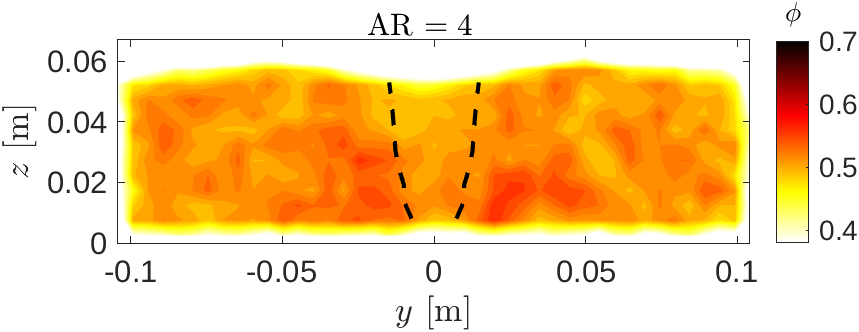}
    \subcaption{}
\end{subfigure}
\begin{subfigure}{0.48\linewidth}
    \includegraphics[trim={0cm 0cm 1.6cm 0cm},clip,width=\linewidth]{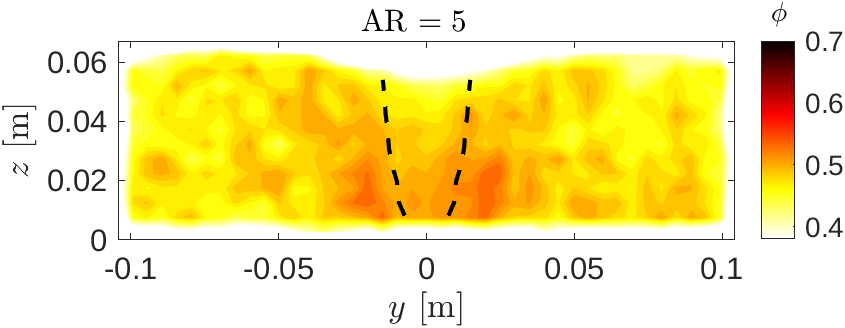}
    \subcaption{}
\end{subfigure}
\begin{subfigure}{0.70\linewidth}
    \includegraphics[trim={1.3cm 9.4cm 1.5cm 3cm},clip,width=\linewidth]{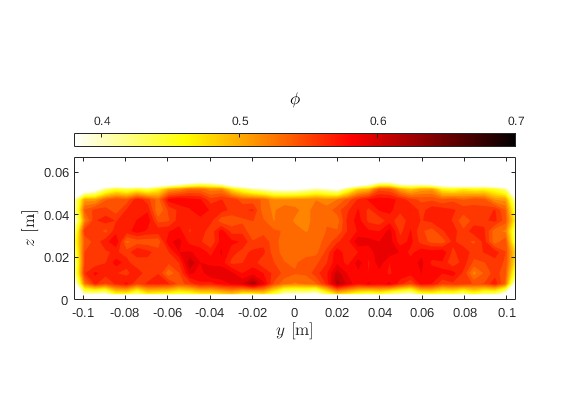}
    \label{fig:subfigC}
\end{subfigure}
\caption{Packing density in the $yz$-plane, averaged over $x$ for different AR. The dashed lines indicate the shear band region calculated from the dimensionless velocity profile in the steady state \cite{ries2007shear,depken2006continuum}. }
\label{fig:PackingDensity}
\end{figure*}

The packing density of particles depends on the particle shape (AR), the friction coefficient, and the shear rate. \autoref{fig:PackingDensity} shows the packing density distribution across the $yz$ plane of LSC, averaged over the $x$ direction, revealing increased filling height for large AR. The dashed lines indicate the shear band location calculated from the dimensionless velocity profile in the steady state \cite{ries2007shear,depken2006continuum}.

For spherical particles, \autoref{fig:PackingDensity}(a), Reynolds dilatancy causes the material to expand under shear leading to low density in the shear band region. Spherical particles rearrange under shear, as they can easily rotate, translate, and slide over each other, therefore, the surface is uniformly flat.

For particles with larger AR, there are two competing effects within the shear band: dilation due to shearing and compaction due to reorientation with the shear direction, resulting in less pronounced dilatation \cite{hosseinpoor2021rheo,marteau2021experimental, hosseinpoor2021rheo}. For AR=5, \autoref{fig:PackingDensity}(d), the compaction effect dominates the Reynolds dilatancy, leading to increased packing density inside and adjacent to the shear band. This combination causes a depression of the surface in the shear band region. 
Our findings are in qualitative agreement experiments by Wegner et al. \cite{wegner2014effects}.

\subsection{Stress analysis}
\subsubsection{Normal stress}
\begin{figure*}[htb!]
\centering
\begin{subfigure}{0.48\linewidth}
    \includegraphics[trim={0cm 0cm 0cm 0cm},clip,width=\linewidth]{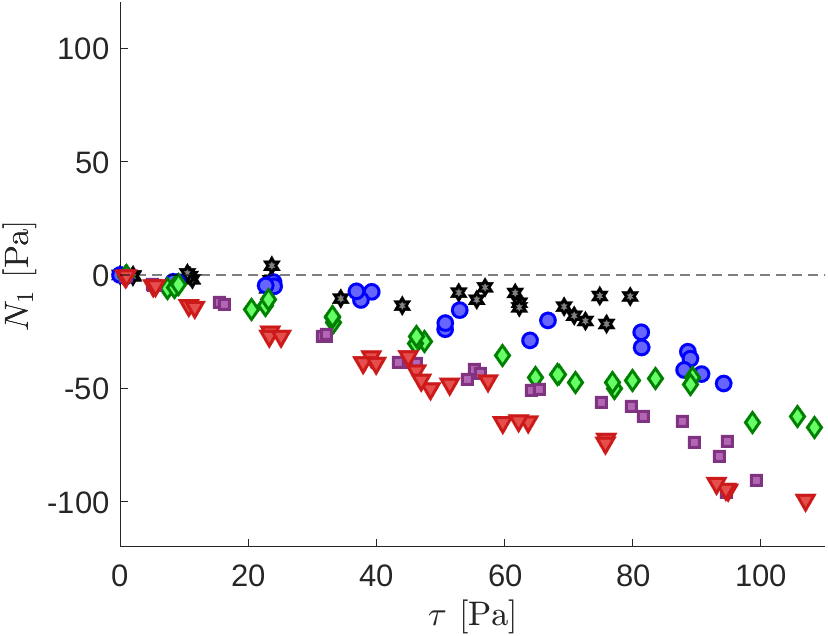}
    \subcaption{}
\end{subfigure}
\hfill
\begin{subfigure}{0.48\linewidth}
    \includegraphics[trim={0cm 0cm 0cm 0cm},clip,width=\linewidth]{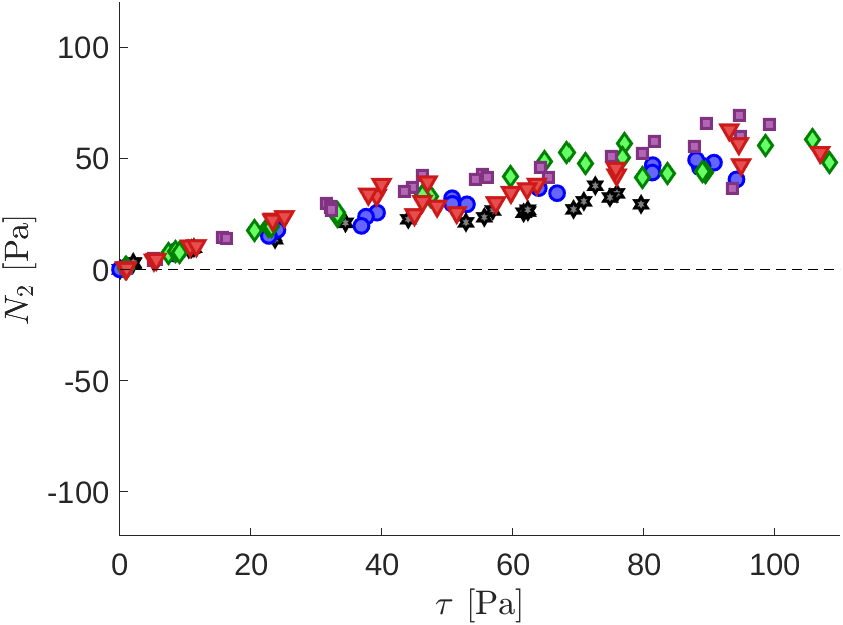}
    \subcaption{}
\end{subfigure}
\begin{subfigure}{0.48\linewidth}
    \includegraphics[trim={0cm 0cm 0cm 0cm},clip,width=\linewidth]{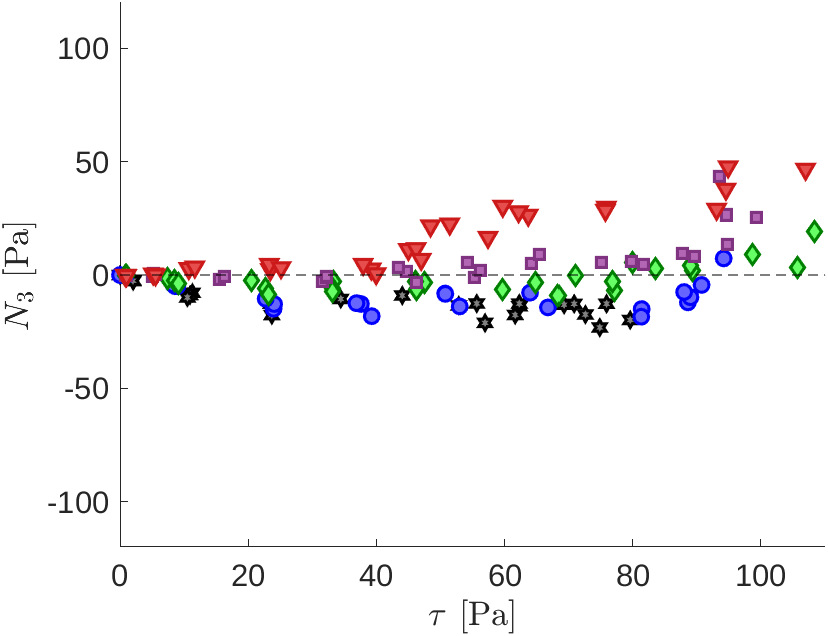}
    \subcaption{}
\end{subfigure}
\hfill
\begin{subfigure}{0.48\linewidth}
    \includegraphics[trim={0cm 0cm 0cm 0cm},clip,width=\linewidth]{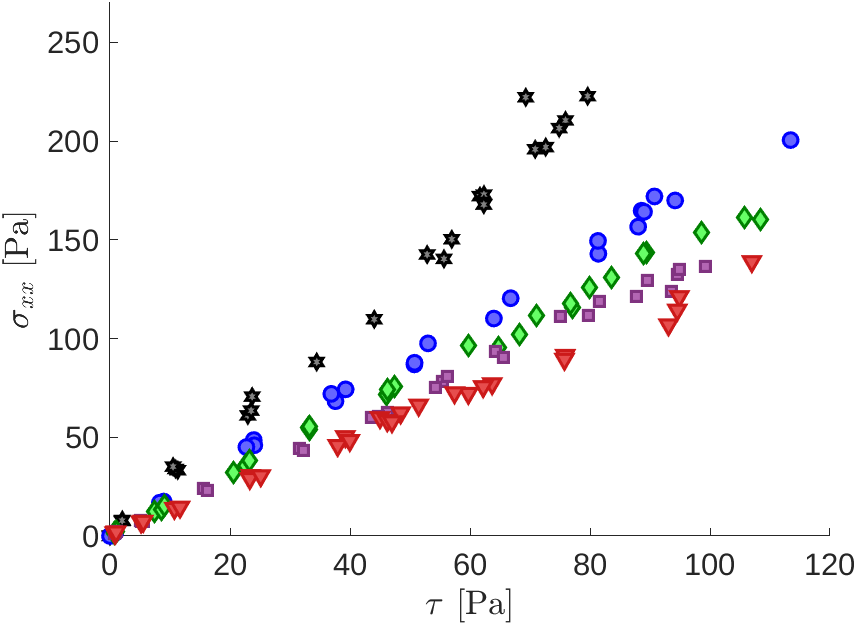}
    \subcaption{}
\end{subfigure}
\begin{subfigure}{0.70\linewidth}
    \includegraphics[trim={1.3cm 12.9cm 1cm 0.5cm},clip,width=\linewidth]{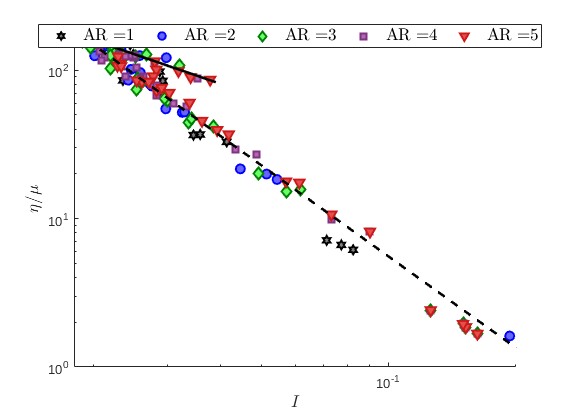}
    \label{fig:subfigD}
\end{subfigure}
\caption{ (a-c) Normal stress differences, $N_1, N_2, N_3$, and (d)  $\sigma_{xx}$  as functions of the shear stress, $\tau$, for various AR.}
\label{fig:stresses_all}
\end{figure*}

Normal stress differences, defined as the differences between the diagonal components of the stress tensor, are fundamental to rheology and, thus, to the analysis of the Weissenberg effect for non-Newtonian fluids \cite{maklad2021review}. In this section, we investigate how these stress differences manifest in granular systems. We examine all the possible combinations of normal stress differences, 
\begin{equation}
    \begin{split}
    N_{1} \equiv &\sigma_{xx} - \sigma_{yy}\\
    N_{2} \equiv &\sigma_{yy} - \sigma_{zz}\\
    N_{3} \equiv &\sigma_{zz} - \sigma_{xx} 
    \end{split}
\end{equation}

as functions of the shear stress 
\begin{equation}
    \tau \equiv \sqrt{\sigma_{xy}^2+\sigma_{xz}^2}\,,
\end{equation}

for various AR, see \autoref{fig:stresses_all}(a-c). 

where $\sigma_{xx}$,
$\sigma_{yy}$, and $\sigma_{zz}$ are the stresses in shear, lateral, and orthogonal directions, respectively.

For all AR, $N_{1}$ shows consistently negative values (see \autoref{fig:stresses_all}(a)), increasing in magnitude with the AR. Conversely, $N_{2}$ maintains positive values for all AR (see \autoref{fig:stresses_all}(b)). Interestingly, $N_{3}$ shifts systematically from negative to positive values with increasing AR from 1 to 5 (see \autoref{fig:stresses_all}(c)). Notably, for spheres, AR=1, all three quantities are nearly zero. 

The behaviors of $N_{1}$ and $N_{3}$ are largely influenced of $\sigma_{xx}$, shown \autoref{fig:stresses_all}(d) as a function of the shear stress. For spheres, AR=1, $\sigma_{xx}$ dominates in magnitude. With increasing AR, the particles align along the streamline, easing the flow of elongated particles in the shear direction. Consequently, this diminishes $\sigma_{xx}$, affecting both $N_{1}$ and $N_{3}$ for larger AR.

For the ordinary Weissenberg effect, where polymer solutions climb along a rotating vertical rod, normal stress differences play a central role \cite{maklad2021review}. The climbing results due to positive $N_{1}$ and negligible $N_{2}$ \cite{maklad2021review}. In contrast, dense suspensions of non-colloidal spheres, where $N_{2}<N_1<0$, exhibit an inverse Weissenberg effect, termed as \textit{rod dipping effect} \cite{maklad2021review}. In our granular system, we observe significant values of $N_{1}$ and $N_{3}$ for large AR, while $N_{1}\approx N_{3}\approx 0$ for spheres, AR=1. 
These stress differences could indicate the granular analogous to the reverse Weissenberg effect (rod dipping effect), where the orientation and shape of particles cause the stress differences. An in-depth analysis is needed to find a more definitive analogy between the classical Weissenberg effect and the observed phenomena in granular matter.

\subsubsection{Shear stress as a function of normal stress}
\begin{figure*}[htb!]
\centering
\begin{subfigure}{0.48\linewidth}
    \includegraphics[trim={0cm 0cm 0cm 0cm},clip,width=\linewidth]{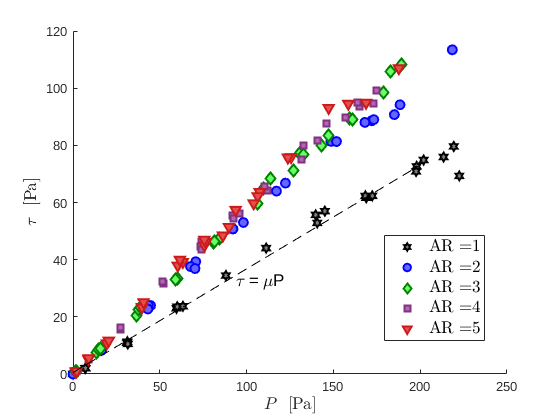}
    \subcaption{}
\end{subfigure}
\hfill
\begin{subfigure}{0.48\linewidth}
    \includegraphics[trim={0cm 0cm 0cm 0cm},clip,width=\linewidth]{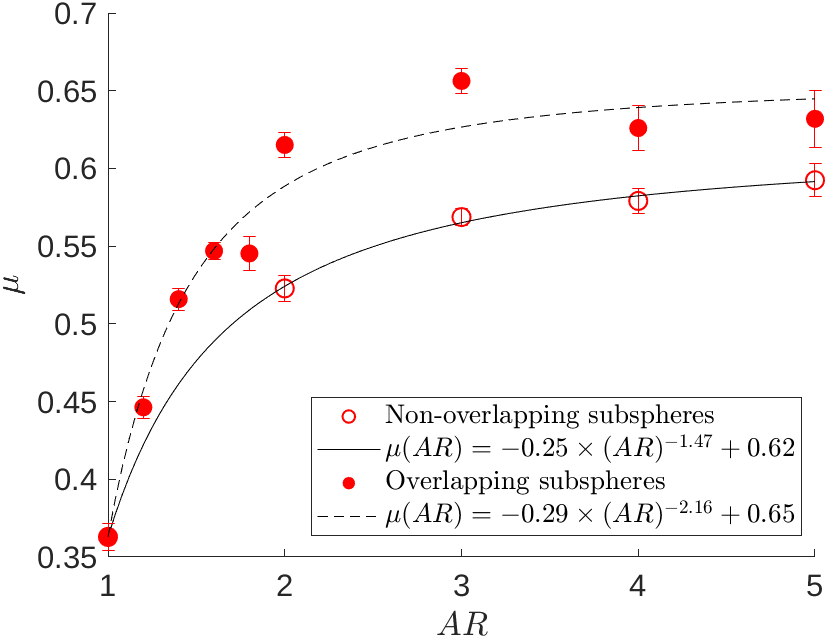}
    \subcaption{}
\end{subfigure}
\caption{(a) Shear stress, $\tau$, as a function of pressure, $P$, for various AR. (b) The macroscopic friction coefficient, $\mu$, increases and saturates with the AR. The lines show fits for both overlapping and non-overlapping subspheres.}
\label{fig:tau_press_mu}
\end{figure*}

The correlation between the shear stress and the normal stress, 
\begin{equation}
P \equiv \frac{1}{3}(\sigma_{xx}+\sigma_{yy}+\sigma_{zz})\,,
\end{equation}
is key to the flow properties of granular materials. \autoref{fig:tau_press_mu}(a) shows a nearly linear increase of the local shear stress as a function of the local normal stress, $\tau=\tau(P)$, where the slope depends on the AR.
This linearity allows for the definition of a macroscopic friction coefficient, $\mu\equiv \tau/P$, that quantifies the material's resistance to motion under shear at a large scale.

Drawn as a function of the AR, \autoref{fig:tau_press_mu}(b), $\mu$ increases and nearly saturates at larger values, in agreement with experimental results by Hidalgo et al. \cite{hidalgo2018rheological}.
The increase of $\mu$ with the AR can be attributed to the additional shear stress required for aligning non-spherical particles with the shear direction. 
For a given AR, particles with overlapping subspheres exhibit a larger friction coefficient, $\mu$, than non-overlapping due to the larger number of individual particle contacts in the granular packing, which increases the shear resistance. 


\subsubsection{Effective viscosity}


We explore the effective viscosity, $\eta \equiv \tau/\dot{\gamma}$ as a function of pressure and strain rate through the inertial number $I$. In granular systems, where the viscosity depends on the confining pressure, it is defined as 
\begin{equation}
I = \frac{\dot{\gamma} d_p}{\sqrt{{P}/{\rho}}}\,.    
\end{equation}

\begin{figure*}[htb!]
\centering
\begin{subfigure}{0.48\linewidth}
    \includegraphics[trim={0cm 0cm 0cm 0cm},clip,width=\linewidth]{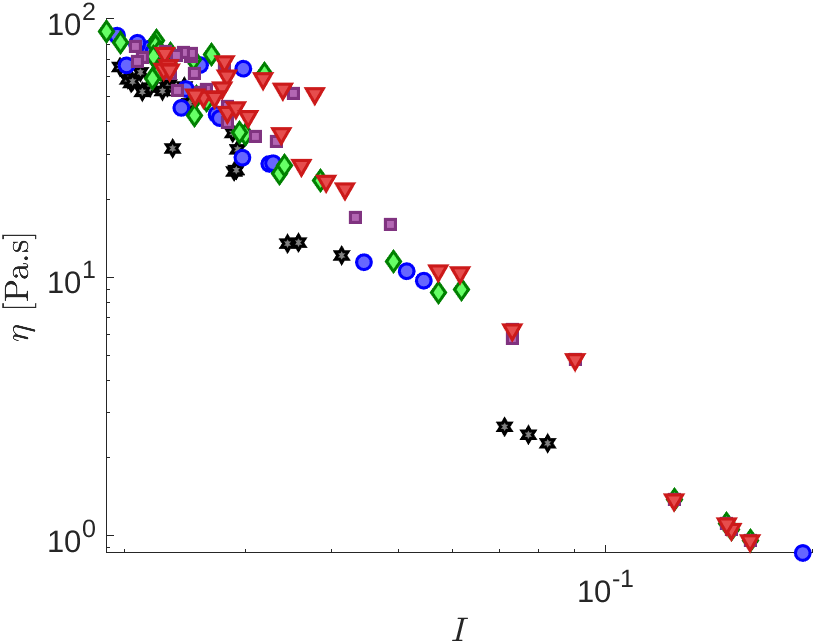}
    \subcaption{}
\end{subfigure}
\hfill
\begin{subfigure}{0.48\linewidth}
    \includegraphics[trim={0cm 0cm 0cm 0cm},clip,width=\linewidth]{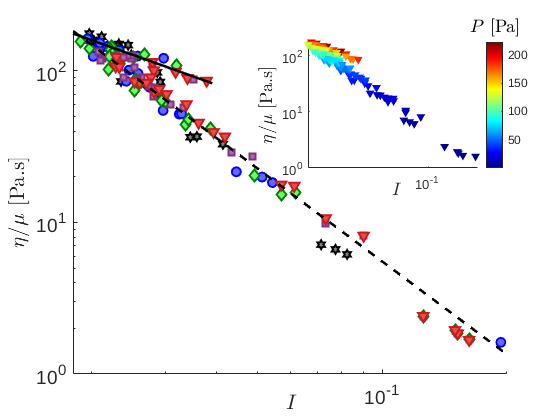}
    \subcaption{}
\end{subfigure}
\begin{subfigure}{0.70\linewidth}
    \includegraphics[trim={1.3cm 12.9cm 1cm 0.5cm},clip,width=\linewidth]{AR_.jpg}
    \label{fig:subfigE}
\end{subfigure}
\caption{(a) Effective viscosity  as a function of inertial number $\eta(I)$ for various values of AR. (b) Scaled effective viscosity $\eta/\mu$ as a function of $I$. Solid and dashed lines show linear fits with slopes -1 and -2, indicating different shear-thinning regimes. The inset shows the pressure-dependent regime of the shear-thinning behavior.}
\label{fig:viscosity}
\end{figure*}

For all values of the AR, the effective viscosity as a function of the inertial number, $\eta(I)$, is a decaying function, see \autoref{fig:viscosity}(a),
indicating shear-thinning behavior. At any given $I$, $\eta$ increases with the AR, that is, granular flows of elongated particles reveal a larger viscosity than flows of spheres. This behavior aligns with the same reasoning explaining the increase of $\mu$ with AR.

\autoref{fig:viscosity}(b) shows the viscosity, scaled by the friction coefficient, $\eta/\mu$, as a function of $I$ to achieve an AR-independent function. The solid and dashed lines are linear fits of two branches with slopes -1 and -2, corresponding to different shear-thinning behaviors, due to the confining pressure (see inset). This result aligns well with previous studies on dry and wet granular flows \cite{luding2007effect,roy2017general}.

\section{Conclusions}

The influence of varying aspect ratios (AR) on particle alignment, packing density, stress, and effective viscosity has been studied. Using a multisphere approach, we modeled particles with both non-overlapping and overlapping subspheres, analyzing how depressions form on the shear band surface when sheared in LSC.

We observed that, with an increase in AR, particles align more readily with the shear direction, mainly within the shear band, indicating higher shear-induced alignment of elongated particles. The global average alignment angle decreases with an increase in AR. This behavior is consistent across multisphere configurations of both non-overlapping and overlapping subspheres.

Spherical particles exhibit dilation within the shear band due to Reynolds dilatancy. However, as the AR increases, particle alignment with the shear direction induces a compaction effect. This leads to a competition between dilatancy and compaction. For higher AR, particularly AR=5, the compaction effect dominates, resulting in denser packing within and adjacent to the shear band.

Rheological analysis reveals that normal stress differences, especially $N_{1}$ and $N_{3}$, are significantly influenced by the AR. Particle alignment eases the flow, reducing the stress in the shear direction $\sigma_{xx}$. The macroscopic friction coefficient $\mu(AR)$ increases with the AR, for both types of multispheres configuration. For a given AR, the overlapping subspheres exhibit a higher friction coefficient, $\mu$, than the non-overlapping ones due to an increased number of particle contacts.

Granular flows exhibit shear-thinning behavior across all AR. Elongated particles, particularly with AR=5, exhibit higher viscosity ($\eta$) than spherical particles for a given inertial number, $I$. Scaling the effective viscosity $\eta$ with the macroscopic friction coefficient $\mu(AR)$, normalizes the influence of AR, showing that scaled effective viscosity $\eta/\mu$ follows a power-law behavior similar to Bagnold's scaling with $I$.


\section{Acknowledgement}
We thank Holger Götz for stimulating discussion. Financial support through
the DFG grant for the Granular Weissenberg Effect (Project No.
PO472/40-1) is acknowledged.




\bibliography{references}

\end{document}